\documentclass[twocolumn,prb,aps,superscriptaddress,showpacs]{revtex4}

\usepackage{dcolumn}
\usepackage{amsmath}
\usepackage{epsfig}

\newlength{\figwidth}
\newlength{\figwidthb}
\begin{document}
\title{Magnetic nature of the 500 meV peak in $\rm La_{2-x}Sr_{x}CuO_4$ observed with resonant inelastic x-ray scattering at the Cu $K$-edge}
\author{D. S. Ellis}
\affiliation{Department of Physics, University of Toronto, Toronto,
Ontario M5S~1A7, Canada}
\author{Jungho Kim}
\affiliation{Department of Physics, University of Toronto, Toronto,
Ontario M5S~1A7, Canada}
\author{J. P. Hill}
\affiliation{Department of Condensed Matter Physics and Materials Science, Brookhaven National Laboratory,
Upton, New York 11973}
\author{S. Wakimoto}
\affiliation{Quantum Beam Science Directorate, Japan Atomic Energy Agency, Tokai, Ibaraki
   319-1195, Japan}
\author{R. J. Birgeneau}
\affiliation{Department of Physics, University of California,
Berkeley, California 94720-7300}
\author{Y. Shvyd'ko}
\affiliation{XOR, Advanced Photon Source, Argonne National
Laboratory, Argonne, Illinois 60439}
\author{D. Casa}
\affiliation{XOR, Advanced Photon Source, Argonne National
Laboratory, Argonne, Illinois 60439}
\author{T. Gog}
\affiliation{XOR, Advanced Photon Source, Argonne National
Laboratory, Argonne, Illinois 60439}
\author{K. Ishii}
\affiliation{Synchrotron Radiation Research Center, Japan Atomic
Energy Agency, Hyogo 679-5148, Japan}
\author{K. Ikeuchi}
\affiliation{Synchrotron Radiation Research Center, Japan Atomic
Energy Agency, Hyogo 679-5148, Japan}
\author{A. Paramekanti}
\affiliation{Department of
Physics, University of Toronto, Toronto, Ontario M5S~1A7, Canada}
\author{Young-June Kim}
\email{yjkim@physics.utoronto.ca}
\affiliation{Department of
Physics, University of Toronto, Toronto, Ontario M5S~1A7, Canada}

\date{\today}

\begin{abstract}
{We present a comprehensive study of the temperature and doping dependence of the 500 meV
peak observed at ${\bf q}=(\pi,0)$ in resonant inelastic x-ray scattering (RIXS) experiments on
$\rm La_2CuO_4$. The intensity of this peak persists above the N\'eel temperature (T$_{N}$=320 K), but decreases gradually with increasing temperature, reaching zero at around T=500 K.
The peak energy decreases with temperature in close quantitative accord with the behavior of the
two-magnon $\rm B_{1g}$ Raman peak in $\rm La_2CuO_4$, and with suitable rescaling,
agrees with the
Raman peak shifts in $\rm EuBa_2Cu_3O_6$ and  $\rm K_2NiF_4$.  The overall dispersion
of this excitation in the Brillouin zone is found to be in agreement with theoretical calculations
for a two-magnon excitation. Upon doping, the peak intensity decreases analogous to the Raman
mode intensity and appears to track the doping dependence of the spin correlation length.
Taken together, these observations strongly suggest that the 500 meV mode is magnetic in
character and is likely a two-magnon excitation.}
\end{abstract}


\maketitle

\section{Introduction}
\label{sect:intro}

Since the discovery of high temperature superconductivity, the cuprate materials
have provided a fertile ground for studying the physics of
electron correlations and quantum magnetism.
Various spectroscopies have been instrumental in revealing the electronic structure of these materials,\cite{Damascelli03,Basov05,Tranquadabook} mostly
focusing on the low energy region below 100~meV, which is of order
of the pseudogap.\cite{Timusk99}  However, one aspect that separates
cuprates from many other condensed matter systems is their intrinsically large energy
scale - a result of the strong hybridization between the in-plane copper and
oxygen orbitals. \cite{Anderson87} This fact has driven recent interest
in spectroscopic studies of the high energy region.  Examples include neutron
scattering studies of the universal spin excitation
spectrum, \cite{Hayden04,Tranquada04,Fujita06,Wilson06,Stock07,Vignolle07} optical studies of the
well-known mid-infrared features,\cite{Sugai00,Lee05, Hwang07, Heumen07} and angle resolved photoemission investigations of high energy kinks.\cite{Graf07,Valla07,Xie07}  Complementing these methods is a relatively new technique, resonant inelastic x-ray scattering (RIXS), which provides momentum resolved information on electronic excitations over a wide energy range.\cite{Kao96,Hill98,Kuiper98,Abbamonte99,Hasan00,Hasan02,Kim02,Kim04d,Ghiringhelli04,Ishii05,Ishii05b,Grenier05,Lu05,Collart06,Wakimoto09,Kotani01}\\


A recent Cu $K$-edge RIXS study of insulating and doped $\rm La_{2-x}Sr_{x}CuO_4$ and
$\rm Nd_{2}CuO_4$ found a new excitation at 500 meV at a momentum
transfer \textbf{q} of $(\pi, 0)$. \cite{Hill08} This mode was found to soften and
lose its spectral weight away from the $(\pi, 0)$ position, and to weaken
in intensity
upon doping holes into the copper-oxygen plane.  The energy scale and
doping dependence of this mode point towards a magnetic origin, and it was argued that the excitation is related
to the two-magnon which has been observed previously with Raman
scattering\cite{Lyons88,Sugai88,Lyons89,Salamon95,Blumberg96,Naeini99,Sugai01,Machtoub05} and optical absorption.\cite{Kim87,Perkins93,Lorenzana95}
While a previous RIXS study at the oxygen $1s$ resonance in $\rm Sr_2CuO_2Cl_2$ showed a weak feature at 0.5 eV that was associated with a two-magnon,\cite{Harada02} this comprehensive Cu $K$-edge study\cite{Hill08} opened an exciting new avenue of studying magnetic excitations with the RIXS technique.  Subsequently, excitations with a two-magnon-like dispersion were observed in both cuprates and NiO with high-resolution $L$-edge RIXS,\cite{Braicovich09,Schlappa09,Ghiringhelli09} and $M$-edge RIXS.\cite{Freelon}  Although excitations in $\rm La_2CuO_4$ at the zone boundary were not presented in this soft x-ray study, the extrapolated
dispersion suggests strong similarity with the hard x-ray RIXS result, giving further credence to the two-magnon interpretation.
Nevertheless, other possibilities, such as a \emph{d-d} crystal field excitation,\cite{Perkins93,Falck94} could not be completely ruled out as possible explanations for the 500 meV mode.\\

In order to further understand this mode, we have carried out comprehensive RIXS studies of the temperature, doping and momentum dependence of the 500 meV peak.  We observe that the RIXS peak becomes significantly weaker in intensity and shifts to lower energy as the temperature is raised
above room temperature, and the peak appears to vanish at around 500 K.  This behavior is very similar to the temperature
dependence of two-magnon Raman
scattering in $\rm YBa_{2}Cu_{3}O_6$ and $\rm EuBa_{2}Cu_{3}O_6$  studied by Knoll et al.\cite{Knoll90}
The momentum dependence of the
peak energy and spectral weight in the insulator is found to be in semi-quantitative agreement when
compared with recent theoretical calculations of the dispersion of two-magnon excitations.
\cite{vandenBrink07,Vernay07,Donkov07,Nagao07,Forte08,Ament09}
Finally, hole doping
causes the peak to shift to lower energy and to lose intensity in a manner very similar to
the behavior of the two-magnon Raman peak.\cite{Sugai88}  Taken together, the above observations strongly suggest that the observed 500 meV peak corresponds to a two-magnon excitation.\\

Despite good agreement in temperature and doping dependence, there remains an important difference between the two-magnon Raman and the RIXS data.  As pointed out in Ref.~\onlinecite{Hill08}, a surprising result was that the 500 meV RIXS peak energy is significantly higher than the $\sim$400 meV low-temperature 2-magnon Raman peak.\cite{Lyons88,Sugai88}  Theory predicts that the 2-magnon Raman peak energy should be renormalized to around 2.75$J$ by strong magnon-magnon interactions,\cite{Lyons88,Canali92,Chubukov95} where $J$ is the effective magnetic superexchange energy including the Oguchi correction.  Such an estimate, for which $J\approx$ 145 meV, would place the 500 meV RIXS peak energy slightly below 3.5$J$.  Since 3.5$J$ corresponds to the peak in non-interacting density of states for \textbf{q}=($\pi$ 0), this implies that magnon-magnon interaction is much less than the \textbf{q}=(0 0) case probed by Raman scattering.\cite{Hill08}  One possible explanation given in Ref.~\onlinecite{Hill08} was that the two spin flips are created on adjacent copper-oxide planes.  Here, we suggest another possibility --- namely, that the interaction between magnons in the same plane is weakened for total momentum ${\bf q}=(\pi,0)$ (as compared to ${\bf q}=(0,0)$ which is explored in Raman scattering). \\


The paper is organized as follows.  Section \ref{sect:experiment} outlines the experimental method.  Section \ref{sect:results}
presents the results for the incident energy and momentum dependence, the temperature dependence of the mid-IR peak and the
effect of doping with holes.  In Section \ref{subsect:disc_of_tdep}, the temperature dependence is discussed in more detail, compared to other materials, and qualitatively described with a simple model.  In section \ref{subsect:disc_of_peakpos}, we discuss a possible reason for the apparently non-interacting value of the observed peak position.  Section \ref{sect:conclude} contains the summary and our conclusions.  The various background subtraction methods
used in analyzing the data are described in detail in Appendix A.\\

\section{Experimental Details}
\label{sect:experiment}

The same $\rm La_2CuO_4$ sample used in Ref.~\onlinecite{Hill08} was used for all of the $\rm La_2CuO_4$ measurements presented here.  We also studied underdoped
$\rm La_{1.93}Sr_{0.07}CuO_4$ to further examine the doping dependence of the 500 meV feature.  This crystal was grown at the University of Toronto by the traveling solvent floating-zone
method, and cut along the $ac$ plane.  After annealing in flowing oxygen for 30 hours at $900\,^{\circ}\mathrm{C}$, the superconducting transition temperature of this sample was 14 K, as determined by the onset temperature of the diamagnetic signal.  The RIXS experiments were carried out at the XOR-IXS 30ID and 9ID beamlines of the Advanced Photon Source, and at the BL11XU beamline at SPring-8, Japan.  The energy calibration was identical for all of the beamlines, with the absorption $K$-edge of a copper foil set to 8980.5~eV.\cite{Deslattes03}  The measurements were all carried out by fixing the incident energy, $E_i$, and scanning the scattered photon energy, $E_f$.  Data are plotted as a function of energy loss $\omega=E_i-E_f$.\\

\begin{figure}
\begin{center}
\epsfig{file=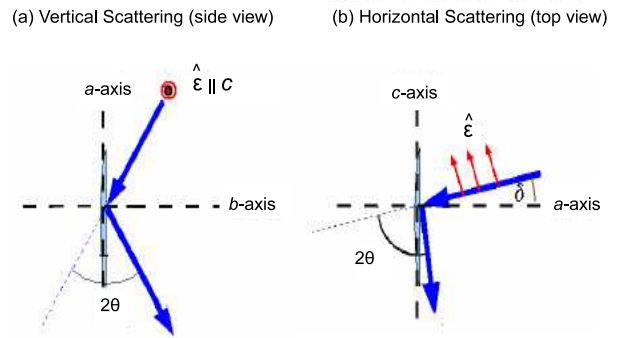,width=3.7in,keepaspectratio}
\end{center}
\caption{The scattering configurations used when scattering in the (a) vertical (sector 9ID experiments) and (b) horizontal plane (at sector 30ID and BL11XU).  The arrows indicate the wavevectors of the incident and scattered beams.  The sample is shown in both cases along the y-axis of the figure (note that the vertical direction in real space corresponds to perpendicular to the incident beam, and parallel to the plane of the page in (a), while it is perpendicular to the plane of the page in (b))  The crystallographic axes are also indicated.  In (a), the incident beam polarization is directed out of the page, parallel to the $c$-axis.  In (b), the incident polarization is in the scattering plane.  The scattering angle 2$\theta$ is set to nearly 90$^{\circ}$, which suppresses the in-plane polarization of the scattered beam.  The angle $\delta$ is about 20$^{\circ}$, so as to have a large  $c$-axis component of the incident polarization.}\label{fig:scatgeom}
\end{figure}

The RIXS measurements at the 30ID beamline were performed using the MERIX spectrometer.  X-rays impinging upon the sample were monochromatized to a bandwidth of 72~meV, using a four-bounce $(+--+)$ monochromator with asymmetrically cut Si(400) crystals.\cite{ToellnerMERIX}.  The beam size on the crystal was reduced to 45(H)$\times$20(V)~$\mu$m$^2$ by two segmented piezoelectric bimorph mirrors, in a Kirkpatrick-Baez configuration (ACCEL Instruments). The photon flux on the sample was $1.2 \times 10^{12}$~ph/s.  A Ge (337) spherical diced analyzer and position sensitive microstrip detector (with channels spaced $125$~$\mu$m) were placed on a Rowland circle of 1~m radius, in order to select $E_f$.  Using a microstrip detector rather than a point-detector allowed measurement of multiple energy channels simultaneously and reduced geometrical broadening of the spectral resolution function.\cite{Huotari05}  As the center channel energy is varied during the energy scan, the same energies are measured multiple times and binned at the end of the scan.  The overall energy resolution for the measurement, with the sample in place, was 120 meV (FWHM) as determined by the width of the elastic peak.\\

The scattering plane was horizontal, with the incident polarization parallel to the scattering plane ($\pi$-polarization) and the scattering angle was set to nearly 90 degrees, thereby suppressing the background from the elastic peak.\cite{Kuiper98,Harada02,Ghiringhelli04,Ishii05,Learmonth09}   In order for the incident beam polarization to have a large $c$-axis component (as in Ref.~\onlinecite{Hill08}) while maintaining a $\sim$90 degree scattering angle, the measurements were carried out at the \textbf{Q}-positions of (3.5 0 6), (3.5 0.25 6), (3.5,0.5,6) and (3 0 8.5) for reduced momentum \textbf{q} of ($\pi$ 0), ($\pi$ $\pi$/2), ($\pi$ $\pi$) and (0 0), respectively.  Here the wavevector change of the scattered beam $\rm \textbf{Q}\equiv\textbf{q}+\textbf{G}$, where \textbf{G} is the reciprocal lattice vector closest to \textbf{Q}.  We use the tetragonal notation, for which the lattice parameter is $\sim$3.78 \AA~along the Cu-O-Cu direction.  A diagram of the scattering geometries is shown in Fig.~\ref{fig:scatgeom}.  Momentum and incident energy dependence measurements were performed at room temperature, while a closed-cycle He refrigerator was used for the temperature-dependence study, which was carried out in a temperature range of 35~K to 500~K.\\

The experimental condition at the 9ID beamline, scattering in the vertical plane, was identical to that of Ref.~\onlinecite{Hill08}, with an overall resolution of 130~meV (FWHM).  All of the measurements at
9ID were carried out near the \textbf{Q}-position of (2.5 0 0), which corresponds to the reduced momentum $\rm \textbf{q}$ of ($\pi$, 0).  \textbf{Q}=(3 0 0) was used for zone center momentum transfer, and (2.5 0.5 0) was used for the \textbf{q}=($\pi$, $\pi$) equivalent position.  The sample was mounted on an Al sample holder and measurements done in the temperature range from 45~K to 300~K.\\

An additional temperature-dependence experiment was carried out at the BL11XU beamline of SPring-8 at the same $\textbf{Q}$.  This instrument had lower resolution (440 meV) and used the same horizontal scattering configuration described above.  Although the lower resolution in this latter setup prevented us from extracting detailed energy parameters, intensity information could be obtained from this experiment, and this is reported in Sec.~\ref{subsect:temp}.  The background subtraction methods used in the various measurements are explained in detail in Appendix A.\\

\section{Experimental Results}
\label{sect:results}

\subsection{Incident Energy and Momentum Dependence}
\label{subsect:Ei and q}

\begin{figure}
\begin{center}
\epsfig{file=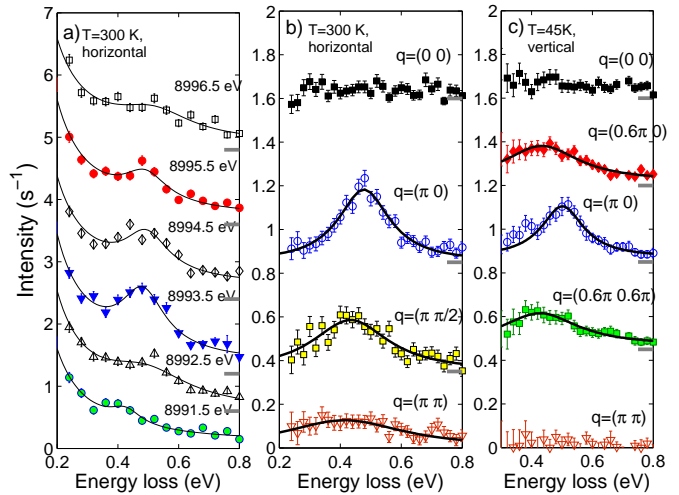,width=3.9in,keepaspectratio}
\end{center}
\caption{(Color Online) (a) Plot of RIXS intensity vs. energy loss, measured at various incident energies at the 30ID beamline (horizontal scattering geometry), for \textbf{q}=($\pi$, 0).  The curves are offset along the y-axis as indicated by the thick grey lines on the right.  The solid lines are
guides to the eye. (b) Background-subtracted intensity is plotted as a function of energy loss, measured at the 30ID beamline at room temperature
for incident energy of 8994 eV, for various momentum transfers. (c) same as (b) but in vertical scattering geometry, from
Ref. \onlinecite{Hill08} (T=45K).  The solids lines are results of fits to Lorentzian lineshapes.} \label{fig:qdep}
\end{figure}

Fig.~\ref{fig:qdep}(a) shows several RIXS spectra for \textbf{q}=($\pi$,0) obtained with different incident photon energies near the absorption $K$-edge in Cu.  The solid lines are guides to the eye.  The incident energy was thereafter set to 8994 eV, around the peak
of the resonance.  The 500 meV excitation resonates at the same photon energy as the higher energy 3.5 eV excitation at this \textbf{q}.\cite{Lu06} The momentum dependence of the energy-loss spectra measured at this incident energy is presented in Fig.~\ref{fig:qdep}(b)-(c).
Fig.~\ref{fig:qdep}(b) shows the room-temperature data taken at the 30ID beamline in the horizontal scattering mode, at various \textbf{q}.  The momentum dependence is in agreement with that observed in Ref.~\onlinecite{Hill08}, which was measured in a vertical
scattering geometry.  These latter data are shown in Fig.~\ref{fig:qdep}(c) for comparison.  In both data sets, the 500 meV peak shows the strongest intensity
at \textbf{q}=($\pi$, 0) and disperses to lower energies away from that maximum.  The peak disappears (or nearly so) at the other two high-symmetry points (0,0) and ($\pi$,$\pi$).\\

\begin{figure}
\begin{center}
\epsfig{file=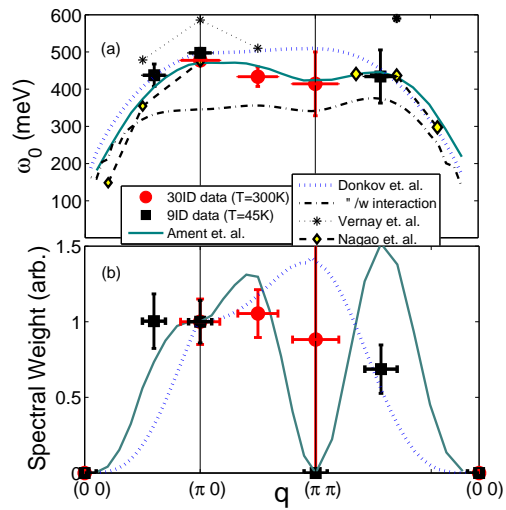,width=3.5in,keepaspectratio}
\end{center}
\caption{(Color Online) (a) Dispersion relation of the two-magnon excitation obtained from two separate measurements as described in the figure legend, and the lines represent theoretical dispersions calculated by Ament et. al.\cite{Ament09}, Donkov et~al.\cite{Donkov07} ($\rm A_{1g}$ - with and without final-state magnon-magnon interaction included in the calculation, as shown by the dash-dotted and vertical dashed lines respectively), Vernay et~al.\cite{Vernay07} ($\rm A_{1g}$), and Nagao et~al.\cite{Nagao07}  The energy scale is normalized
by the magnetic exchange constant $J$ ($\approx$145 meV).  Here we suppose that 3.5$J$ corresponds to 500 meV.  The horizontal error bars span the \textbf{q}-resolution.  (b) Spectral weight of the background-subtracted data, defined as the product
of the Lorentzian peak intensity and linewidths; this does not include constant background, so that the spectral weight at zone center is zero.   The weights are normalized by the weight at \textbf{q}=($\pi$, 0).  Same legend as in (a), two available theoretical calculations by Ament et al. and Donkov et al. are plotted.} \label{fig:mag_disp}
\end{figure}

The main difference between the results from the horizontal and vertical geometry measurements is that there appears to be some intensity at \textbf{q}=($\pi$,$\pi$) in the horizontal scattering data set, Fig.~\ref{fig:qdep}(b), whereas this was negligible in the vertical scattering experiment.  The discrepancy might be due to a non-resonant scattering contribution which would remain after the energy-gain side background subtraction procedure, as discussed in Appendix A.  Polarization could be another possible explanation, since in the horizontal scattering case, the incoming polarization has a non-zero $a$-axis component, and the outgoing photon has a reduced $c$-axis component, when compared to the vertical scattering geometry.  However, a more likely reason for this discrepancy is the momentum resolution effect.  The momentum resolution at \textbf{q}=($\pi$,$\pi$) in the horizontal scattering experiment extended $\sim$20\% towards ($\pi$ 0), while the \textbf{q}-resolution in vertical scattering extended only $\sim$6\% towards ($\pi$ 0) (the resolution is shown as the horizontal error bars in Fig.~\ref{fig:mag_disp}).  The lower momentum resolution means that the data obtained at nominal \textbf{q}=($\pi$,$\pi$) contains significant contributions from the neighboring \textbf{q}, giving rise to non-zero intensity.  When the intensity is a rapidly varying function of \textbf{q}, such as may be the case near \textbf{q}=($\pi$,$\pi$),\cite{Forte08} full resolution deconvolution would be required to describe our intensity observations quantitatively.  Because of the higher \textbf{q}-resolution, and a background-subtraction which removes non-resonant scattering contributions, we are more confident in the vertical scattering result at \textbf{q}=($\pi$,$\pi$).\\

The measured spectra were each fit to a Lorentzian plus constant background, shown by the solid lines in Fig.~\ref{fig:qdep}(b)-(c).  The resultant peak positions are
plotted in Fig.~\ref{fig:mag_disp}(a).
In Fig.~\ref{fig:mag_disp}(b), we plot the spectral weight, calculated by multiplying the Lorentzian
peak width and intensity parameters together.  The spectral weights in Fig.~\ref{fig:mag_disp}(b) are scaled such that the spectral
weight at \textbf{q}=($\pi$, 0) is set to unity.  Since there is no observable Lorentzian peak at \textbf{q}=(0 0), the spectral weight there was set to zero (and no error-bars were assigned to those zero points). \\ 

Motivated by the original observation of the 500 meV mode, there have been a
number of recent theoretical calculations \cite{vandenBrink07,Vernay07,Donkov07,Nagao07,Forte08,Ament09}
for the dispersion of the two-magnon peak in a Heisenberg antiferromagnet.
Nagao and Igarashi\cite{Nagao07} and the van den Brink group\cite{vandenBrink07,Forte08,Ament09}
explicitly included a core-hole interaction in their cross-section in order to account for the effects of the
intermediate state in the RIXS process.  Donkov et~al.\cite{Donkov07} and Vernay et~al.\cite{Vernay07}
calculated a generalized momentum-dependent Raman response for both $\rm A_{1g}$ and
$\rm B_{1g}$ scattering configuratations, without including core-hole effects.  A comparison of the theoretical
calculations and the experimental data is provided in Fig.~\ref{fig:mag_disp}.
Since the lineshapes
of the calculated spectra are typically not Lorentzian, we chose to compare the first moment of the
computed spectra, which is likely to be a more robust feature of the theoretical calculations.
For clarity, we include only the latest calculation from the van den Brink group,\cite{Ament09} and
only the $\rm A_{1g}$ geometry results from Refs.~\onlinecite{Vernay07} and \onlinecite{Donkov07}.  Although we cannot make
a direct link between the scattering geometries of RIXS and Raman spectroscopies due to the fundamental differences in the processes,
the $\rm B_{1g}$ mode has high intensity and energy at \textbf{q}=0, which we do not observe in any our RIXS data, so we compare only to the $\rm A_{1g}$ calculations.
The theoretical energies, originally given in units
of $J$, are multiplied by the value of $J$ to compare with the experiment.  The value of $J$ used here is 143~meV, so as to give 3.5$J$=500 meV.
Given the considerable differences between the Raman scattering and RIXS processes, one should compare these results with caution.\\

Fig.~\ref{fig:mag_disp}(a) shows reasonable agreement between the various theoretical dispersion calculations and
the experiment, especially the RIXS-specific calculations.  Discrepancies include the fact that when interactions are included in the calculation of
Donkov and Chubukov,\cite{Donkov07}
the predicted energies are systematically lower than the observed values.  As pointed out by these authors, it is possible that the random-phase-approximation-type analysis overestimated the effect of the interactions.
The calculated spectral weights are plotted for two of the theoretical calculations\cite{Donkov07,Ament09}
in Fig.~\ref{fig:mag_disp}(b).
A main point of agreement between these calculations and the experiment is the suppression
of intensity at the zone center.  However, for \textbf{q}=($\pi$ $\pi$) there is a large difference between these
calculations.  This also happens to be the \textbf{q} position where the horizontal and
vertical scattering measurements disagree, but the experimental considerations discussed above would support the zero intensity at \textbf{q}=($\pi$ $\pi$) result.\\

Besides the suppression of spectral weight near zone center and relative flattness near \textbf{q}=($\pi$ 0), there is no obvious overlap in spectral weight between the theories and experiment.  This is not very surprising considering the many factors that may influence spectral weight as a function of momentum, limiting the spectral weight's use as a detailed comparison with theory. In contrast with this uncertainty, the energy dispersion along ($\pi$ 0)-($\pi$ $\pi$), as seen by the ($\pi$ 0) point and the new ($\pi$ $\pi$/2) point in Fig.~\ref{fig:mag_disp}(a), is quite consistent with the recent RIXS calculations, as are the energies at the other \textbf{q}-positions.  On the basis of this result it would seem that the energy dispersion calculations incorporating the core-hole potential\cite{vandenBrink07,Forte08,Ament09} agree closely with our data, while the pure Raman calculations\cite{Donkov07,Vernay07} agree somewhat less consistently.\\



\subsection{Temperature dependence}
\label{subsect:temp}

\begin{figure}
\begin{center}
\epsfig{file=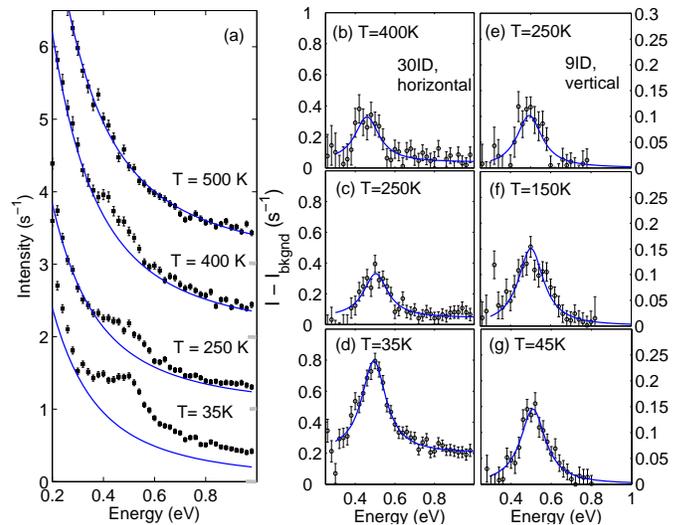,width=3.7in,keepaspectratio}
\end{center}
\caption{(Color Online) (a) Temperature dependence of ($\pi$, 0) RIXS spectra (black squares) of $\rm La_2CuO_4$ obtained at 30ID.  The background (thin line) was determined by scaling a Lorentzian fit of the 500~K data by the
elastic line intensity at each temperature, as discussed in the text.  The spectra are displaced along the y-axis for clarity.  The thick grey ticks on the right side of the graph indicate zero levels.  (b)-(d) Background-subtracted data and fits for T=400~K, 250~K,
and 35~K, respectively.  (e)-(f) The spectra measured at 9ID for temperatures of 250~K, 150~K and 45~K, respectively.  The background
 for the 9ID data was obtained from off-resonance data as described in Section \ref{sect:experiment}.} \label{fig:Tempdata}
\end{figure}

The 500 meV peak at $(\pi, 0)$ was measured at various temperatures.  The spectra obtained at 30ID are shown in Fig.~\ref{fig:Tempdata}(a), with a clear 500 meV peak at low temperatures.  As the temperature is increased, the peak becomes weaker, and at T=500~K there appears to be
little trace of the peak left. The data at 500~K can be well described as the tail of a Lorentzian centered at zero
energy, and therefore we use this as our background.  At each temperature, the Lorentzian fit to the T=500~K data was scaled to match the energy loss
$\omega$=0 intensity, and used as a background (shown as solid lines in
Fig.~\ref{fig:Tempdata}(a)).  Note that for T=35~K, the background
calculated in this way is below the experimentally observed signal, which might imply considerable constant intensity in addition to the main peak, as seen in the figure.  We do not understand the source of this additional intensity at present, and here we only focus on the well-defined peak.  The background-subtracted data are
shown in Fig.~\ref{fig:Tempdata} (b)-(d). The solid lines are results of fits to a simple Lorentzian
function, with an additional constant background for the T=35~K data.  Given the error bars on the data, and that
the peak width is not much larger
than the instrumental resolution, we could not reliably extract the temperature dependence of the peak widths from the
data. The fits were therefore carried out simultaneously for all curves keeping the peak width the same
for all temperatures.  The resulting width was found to be 166$\pm$10 meV.  A rapid suppression of the intensity as
temperature is increased is clearly observed.
Measurements of the temperature dependence were also carried out at 9ID and at BL11XU.  In Fig.~\ref{fig:Tempdata}(e)-(g), we plot
the background-subtracted data obtained at 9ID.  Note
that in this case, the off-resonance scans were used as background scans as described in Appendix A.  The solid lines
are curves of Lorentzian fits, with widths of 153$\pm$10 meV which is similar to the 30ID result.   \\

\begin{figure}
\begin{center}
\epsfig{file=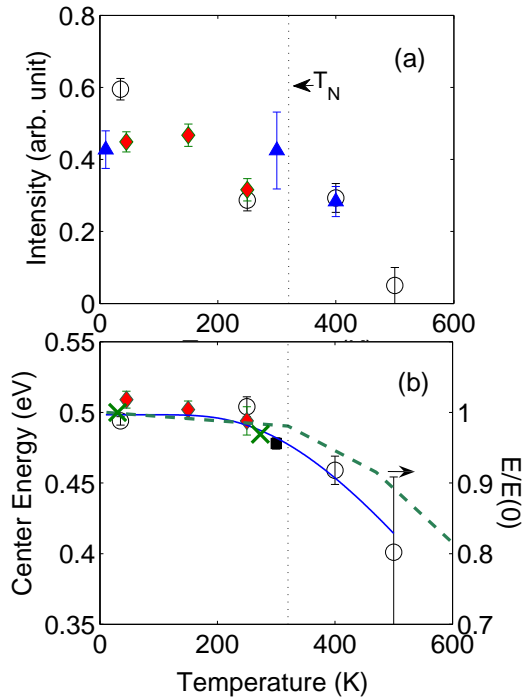,width=4.5in,keepaspectratio}
\end{center}
\caption{(Color Online) Temperature dependence of the Lorentzian fit parameters for the RIXS peaks measured at \textbf{q}=($\pi$, 0).
Shown are the fit parameters (a) intensity and (b) excitation energy.  Different symbols correspond to data taken with different experimental
conditions: horizontal scattering at 30ID (Open circles and filled square.  The filled square corresponds to a seperate beamtime);
Vertical scattering at 9ID (filled diamonds); and horizontal scattering at BL11XU (filled triangle).  The solid line is a result of the fit to Eq.~(\ref{eq:shift}), with fit parameters $E_0$=0.50 eV, $\gamma$=1.7, and $\Omega$=0.1 eV.  The dotted vertical line indicates the N\'eel temperature for $\rm La_2CuO_4$.  The dashed curve, which is referred to the right-hand scale,
is the temperature dependence of the two-magnon Raman peak position of $\rm EuBa_{2}Cu_{3}O_6$ obtained from the peaks of the fitted curves in
Ref.~\onlinecite{Knoll90}.  On the same scale, the crosses are the two-magnon Raman peak position for $\rm La_2CuO_4$ as measured in Ref.~\onlinecite{Sugai88}.}
\label{fig:Lor_fit_params}
\end{figure}

Figure~\ref{fig:Lor_fit_params} shows the parameters obtained from the Lorentzian fitting of the data as a
function of temperature.  Note that the intensity at 500 K comes from the residual intensity between 0.4 and 0.5 eV, after subtracting the fit curve, but is zero within error bar.  Also shown is the intensity data obtained at BL11XU. The Lorentzian linewidth of $\sim$160 meV (FWHM) is
close to being resolution-limited in all of the scans, and thus it was not
possible for us to detect any temperature dependence to the
line broadening, although we cannot rule it out.
The intensity data from the 9ID and BL11XU experiments, shown in Fig.~\ref{fig:Lor_fit_params}(a), are scaled to
approximately match the 30ID intensities at the temperatures
they share in common. The scatter in the intensity data is fairly large but
the overall trend clearly shows the intensity decreasing, reaching
nearly zero at 500~K.  The peak position, plotted in Fig.~\ref{fig:Lor_fit_params}(b), shifts to lower frequency as temperature increases. The solid line comes from fitting to a phenomenological equation, presuming that a two-magnon mode would interact with ambient Bose particles (magnons or phonons),
\begin{equation}   E(T) = E_0\cdot(1-\gamma\cdot n(\Omega,T))  \label{eq:shift}\end{equation}
where  $E_0$ is the peak energy at T=0 K.  $\gamma$ and $\Omega$ are fitting parameters, and
\emph{n}($\Omega$,T)=$\frac{1}{exp(\Omega/k_BT)-1}$ is the Bose factor of characteristic energy $\Omega$.  The peak position at room temperature is about 5\% smaller than at 35~K.  The magnitude of the peak shift lies on the same curve as the shift of the two-magnon Raman scattering peak of $\rm La_{2}CuO_4$ observed by Sugai, et. al.\cite{Sugai88}, who report a 3\% downshift
for T=273 K, shown in Fig.~\ref{fig:Lor_fit_params}(b).  Note that this is also consistent with the Raman peak shifts more recently measured by Sugai and Hayamizu,\cite{Sugai01} (not shown in Fig.~\ref{fig:Lor_fit_params}(b) to avoid over-crowding the graph).  A neutron scattering study of the 1-magnon dispersion in $\rm La_{2}CuO_4$ has
reported that the zone boundary magnon energy at T=300~K is about 4\% lower than its value at T=10~K,\cite{Coldea01} which is also consistent with the current result if we assume that the two-magnon energy is proportional to the one-magnon energy. Thus, for room temperature and below, the shift in the 500 meV peak agrees quantitatively with the two-magnon Raman result and also scales with the zone boundary 1-magnon energy.\\

For higher temperatures, similar temperature dependence of the excitation energy has been observed for the two-magnon Raman
spectra of $\rm EuBa_{2}Cu_{3}O_6$.\cite{Knoll90}  The excitation energy of $\rm EuBa_{2}Cu_{3}O_6$ two-magnons obtained from the maxima of the fitted curves in the Fig.~1 of Ref. \onlinecite{Knoll90} are plotted in Fig.~\ref{fig:Lor_fit_params}(b), and compared with that of $\rm La_2CuO_4$.  There is similarity in
the temperature dependence of the two cuprates, although the peak position decreases more rapidly with temperature in $\rm La_2CuO_4$.  The temperature dependence will be discussed in more detail in Sect.~\ref{subsect:disc_of_tdep}.

\subsection{Doping Dependence}
\label{subsect:dope}

\begin{figure}
\begin{center}
\epsfig{file=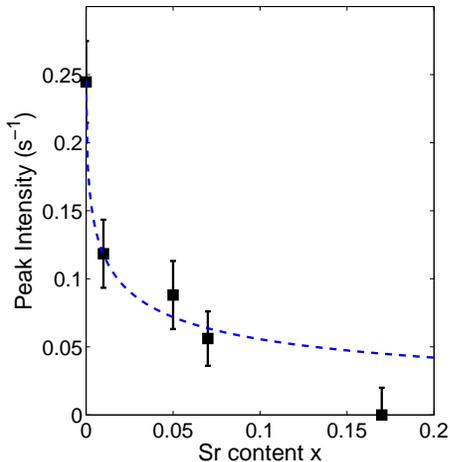,width=3.5in,keepaspectratio}
\end{center}
\caption{Peak intensity of the 500 meV peak vs. doping $x$ for $\rm La_{2-x}Sr_{x}CuO_4$.  The peak intensities were determined from Lorentzian fits of the spectra from Ref. \onlinecite{Hill08} and in Fig.~\ref{fig:07_back_subtract}.  Between different samples, the spectra were normalized by the intensity of the elastic line at the same \textbf{Q} position and incident energy.  The dashed line results from a fit to the form $I(x)=I_0/(1+r \sqrt{x})$, with $r\sim$10.}
\label{fig:I_vs_x}
\end{figure}

In addition to the samples measured in Ref.~\onlinecite{Hill08}, the RIXS spectrum of a $\rm La_{2-x}Sr_{x}CuO_4$ sample of $x$=0.07 was also measured at 9ID using the vertical scattering geometry (the spectrum is shown in Appendix A).  Combining data from Ref. \onlinecite{Hill08} and the current work, the doping dependence of the 500 meV peak intensity is plotted in Fig.~\ref{fig:I_vs_x}. For this figure, the peak intensities were normalized from sample to sample by the elastic ($\omega=0$) peak intensity.  The error bars in relative intensity are then large due to the elastic peak intensity variations from sample to sample, depending on the crystal quality.  The general trend, however, is that as $x$ increases, the spectral feature intensity decreases, but the peak feature survives up to at least $x$=0.07, well into the superconducting phase.  For $x$=0.17, the peak appears to be either absent or highly damped.\\

It is well known that magnetic correlations are suppressed upon hole-doping in the cuprates, so that
this observed doping dependence of the intensity suggests that the excitation has a magnetic character.~\cite{Hill08}
The existence of the peak for doping values $x=0.07$ shows that this excitation survives even if long-range magnetic order is absent, so long as significant short-range magnetic correlations are present (which is well known to be the case in the superconducting state of LSCO\cite{Wakimoto04,Lipscombe07}). This is consistent with our observation that the mode also survives above the N\'eel temperature in the undoped parent compound.  The doping dependence of the RIXS intensity is quite similar to
the doping dependence of the two-magnon Raman spectra in Ref.~\onlinecite{Sugai88}, in which there is a strong drop in intensity occurring between $x$=0 and $x$=0.01.  The peak is still visible at $x$=0.07, but no intensity appears on the same scale for $x$=0.12 (note that $x$ in our notation is twice that of Ref.~\onlinecite{Sugai88}).\\

 Assuming that the 500 meV mode is two-magnon in origin, the following physical argument sheds some light on the
doping dependence of the intensity. The instantaneous spin-spin correlation length $\xi$, which is
a measure of magnetic correlations, depends on doping as approximately $\xi \sim a/\sqrt{x}$ over a range of dopings, where $a$ is the lattice constant.\cite{Thurston89b}  While there is a certain inverse lifetime, $\Gamma_0$, to two-magnon excitations even in the insulating compound (energy resolution also contributes to the nominal width), we expect the finite magnetic correlation length in the doped system to lead to an additional scattering rate for two-magnon excitations.  This scattering rate may be calculated by $\Gamma_x \sim {\it v}/\xi$, where ${\it v}$ is the characteristic velocity associated with magnetic excitations. Assuming that the spectral weight of the 500 meV mode is not a strong function of doping for small $x$, the peak intensity is expected to scale as $1/(\Gamma_0 + \Gamma_x) \sim 1/(1 + r\sqrt{x})$ where $r \sim \frac{v}{a \Gamma_0}$.  However, we caution that using the correlation length determined in Ref.~\onlinecite{Thurston89b} for low energies should not generally apply for these high-energy excitations.  Nevertheless, we find that this functional form, $I(x)=I_0/(1+r\sqrt{x})$ fits the data quite well up to $x\sim 0.07$ as shown in Fig.~\ref{fig:I_vs_x}. Setting $\Gamma_0 =$~150 meV consistent with the observed Lorentzian linewidth of the mode and ${\it v} \sim J a$ as a rough measure of characteristic velocity, we find a value for $r$ which is within order of magnitude of the value obtained from the best fit to the data, lending support to the above argument.

\section{Discussion}
\label{sect:discuss}

\subsection{Temperature dependence}
\label{subsect:disc_of_tdep}

Magnon excitations and magnon-magnon interactions in two-dimensional (2D)
S=1/2 Heisenberg antiferromagnets have been extensively investigated over many years, \cite{Grempel88,Singh89,Kopietz90,Tyc90,Canali92,Makivic92} especially after the discovery of high temperature superconductivity.
Theoretical calculations have shown that the short wavelength magnon near the zone boundary is
well defined even above the N\'eel temperature \cite{Tyc90}--
that is, the damping due to magnon-magnon interactions $\Gamma(T)$ is much smaller than the magnon energy
as long as q$\xi \gg 1$.  A number of studies have looked into the damping of two-magnon excitations due to magnon-magnon
interactions.\cite{Bacci91,Canali92,Saenger95,Manning95,Sandvik98} The consensus
from these studies is that the magnon-magnon interactions alone
cannot explain the observed broad linewidth of the two-magnon Raman scattering or the temperature dependence of the
intensity. In order to explain the observed linewidth in their two-magnon Raman scattering results on
$\rm EuBa_{2}Cu_{3}O_6$, Knoll et. al. suggested that magnon-phonon interaction\cite{Akhiezer46,Cottam74} is an additional source of damping.\cite{Knoll90}
Such damping from phonons is expected to be a larger effect in cuprates than in compounds such as $\rm K_2NiF_4$, since the magnetic energy scale in the cuprates is much larger than typical phonon energies, resulting in smaller thermal population of magnons compared to that of phonons.\cite{Knoll90,Saenger95}  In $\rm La_2CuO_4$, the damping may even be more pronounced due to a rotational phonon mode which softens towards the structural transition temperature at around T=500 K.\cite{Birgeneau87,Boni88}  This damping may well be responsible for the disappearance of the peak at around the same temperature, although further temperature dependence studies would be required to test this.\\

With the energy resolution used in this study, linewidth broadening is difficult to detect.  Instead, we focus on the shift in the peak energy.  Temperature-induced softening of the two-magnon energy in $S\!=\!1$ compounds such as $\rm K_2NiF_4$ has been quantitatively explained by magnon-magnon interaction theory. Keffer and Loudon\cite{Keffer61}, Bloch\cite{Bloch63}, and Davies et. al.\cite{Davies71, Chinn71}, calculated the shifts in the magnon energies $\Omega_\textbf{k}(T)$
to be proportional to the number of magnons determined by the Bose distribution.
For a Heisenberg antiferromagnet, the one-magnon dispersion gets renormalized as
\begin{eqnarray}   \Omega_\textbf{k}(T) &=& \alpha(T)\Omega_\textbf{k}   \label{eq:Omega_k}\\
\alpha(T) &=& \alpha(0) -\frac{1}{JzS^2}\frac{1}{N}\displaystyle\sum_{\mathbf{q}}\frac{\Omega_\mathbf{q}}{e^{\Omega_\mathbf{q}\alpha(T)/k_BT}-1}  \label{eq:alpha}\end{eqnarray}
where $J$ is the magnetic exchange constant, $z$ is the number of nearest neighbors, $S$ is the spin, and
the renormalization factor $\alpha(T)$ needs to be determined self-consistently.  The constant $\alpha(0)$ incorporates the
Oguchi \cite{Oguchi60} correction to the spin wave velocity at zero temperature.
(We note that if the sum in Eq.~(\ref{eq:alpha}) is dominated by a single momentum, it reduces to the form of
Eq.~(\ref{eq:shift}) which motivated our earlier fit in Fig.~\ref{fig:Lor_fit_params}).
%
%
It is reasonable to assume that the two-magnon excitations get renormalized in a similar fashion.
In order to compare different materials, it is then more useful to cast the expression
for the energy-shift $\Delta E\equiv E_0-E(T)$ of the two-magnon excitation, normalized to its low-temperature value $E_0$,
in the form
\begin{equation}
\frac{\Delta E}{E_0} \approx  \frac{1}{S}\cdot f(\frac{T}{SJz}) \label{eq:shift_norm}
\end{equation}
where $f(\tilde{T})$ is a scaling function obtained from Eq.~(\ref{eq:alpha}) with its argument being the
scaled temperature $\tilde{T}=T/(SJz)$.  Here $S$ is the spin quantum number, $z$ is the number of nearest neighbors, and $J$ is the magnetic exchange energy. Such a scaling ansatz is valid when the interaction corrections to
$\alpha(T)$ are small, so that we can set $\alpha(T)\approx 1$ in the Bose function in Eq.~(\ref{eq:alpha})
and assume
$\Omega_{\textbf{k}} \propto J z S$.
To leading order, Eq.~(\ref{eq:shift_norm}) determines the normalized energy shift in terms of the basic
magnetic parameters of a material. The prefactor of the energy shift scales as $1/S$ since this governs the
strength of magnon-magnon interactions.\cite{Oguchi60}  Such spin-dependence is evident from the observed two-magnon Raman energies: for spin-1 $\rm K_2NiF_4$, the two-magnon energy is renormalized from 8$J$ to 6.8$J$, which is a 15\% decrease from its theoretical value without spin-spin interaction, while for spin-1/2 cuprates, the renormalization from 4$J$ to 2.7$J$ represents a 32.5\% decrease, or about twice as much.\\

\begin{figure}
\begin{center}
\epsfig{file=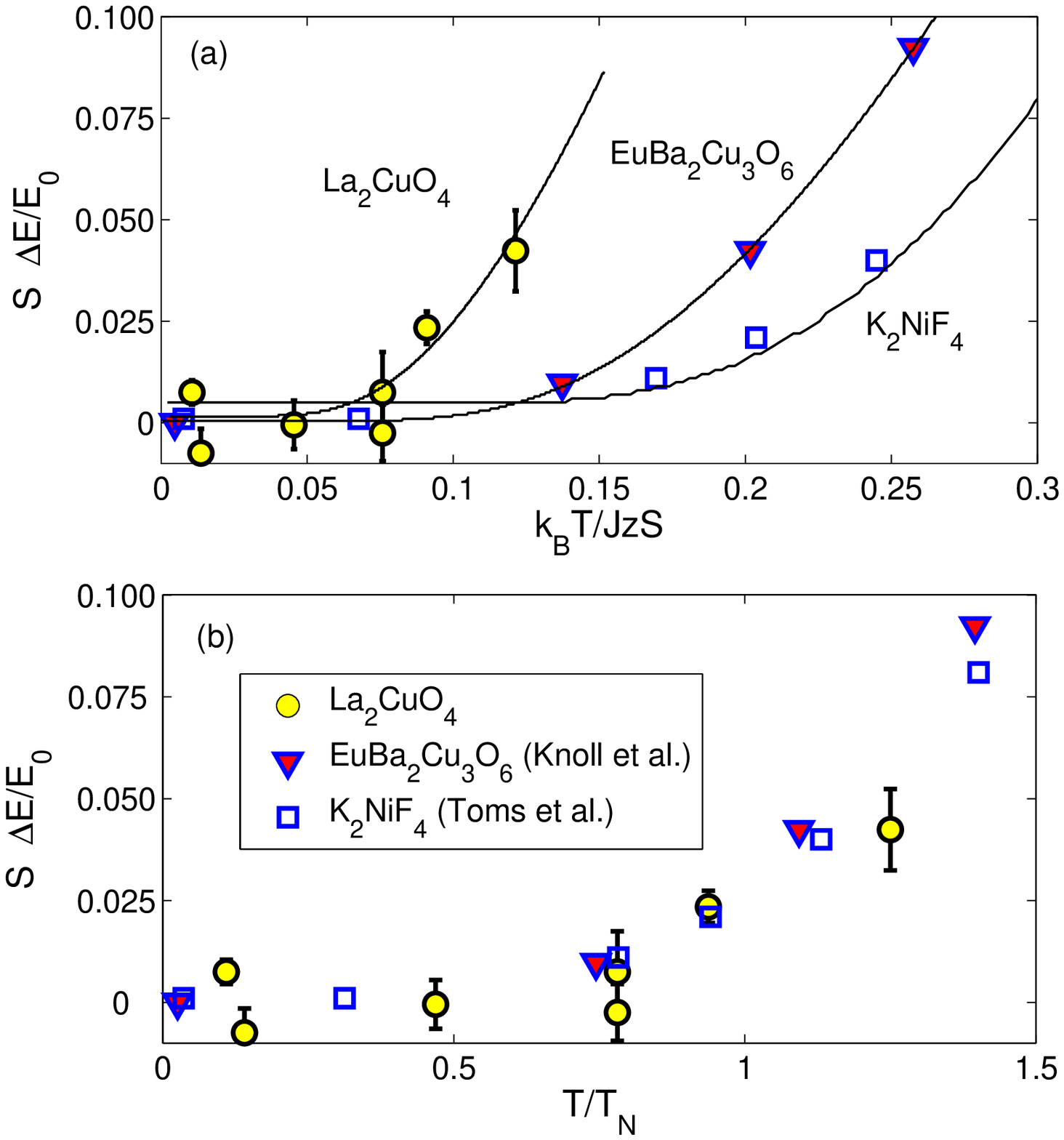,width=5in,keepaspectratio}
\end{center}
\caption{(Color Online) Comparison of scaled peak energy shift vs. scaled temperature in
$\rm La_2CuO_4$ as measured by RIXS in this study (circles),
and two-magnon Raman scattering in  $\rm EuBa_2Cu_3O_6$ (triangles)
obtained from the peaks of the fit curves in Ref.~\onlinecite{Knoll90}, and $\rm K_2NiF_4$ (unfilled squares) from Ref.~\onlinecite{Toms74}. (a) The peak energy shift is normalized by the low-temperature energy, and further multiplied by spin $S$, as discussed in the text.  The temperature scales are normalized by $JSz$.  $J$ of 143 meV is used for $\rm La_2CuO_4$, 100 meV for $\rm EuBa_2Cu_3O_6$ \cite{Knoll90} and 9.7 meV for $\rm K_2NiF_4$ \cite{Toms74} respectively.  The lines result from fits to Eq.~(\ref{eq:shift}).  (b)  Same as (a), except normalizing the temperature to the N$\acute{e}$el temperature $T_N$, which is 320K for $\rm La_2CuO_4$, 430K for $\rm EuBa_2Cu_3O_6$\cite{Knoll90} and 97K for $\rm K_2NiF_4$ \cite{Fleury70} (in Ref. \onlinecite{Toms74} the temperature is already normalized to $T_N$) .}
\label{fig:compare_materials}
\end{figure}

We now check if the observed energy shifts can be explained by the spin-spin interaction theory by comparing the shifts in different 2D Heisenberg antiferromagnet compounds.  The scaling function in Eq.~(\ref{eq:shift_norm}) is plotted in Fig.~\ref{fig:compare_materials}(a) for our RIXS data in $\rm La_2CuO_4$, and the two-magnon Raman data in $\rm EuBa_{2}Cu_{3}O_6$\cite{Knoll90} and those of $\rm K_2NiF_4$.\cite{Toms74}  The peak shifts in the latter material were proven to follow the spin-spin interaction theory, at least up to $\tilde{T}$=0.3,\cite{Chinn71} so this curve may be considered as the theoretical prediction.  The clear non-overlap of the curves in Fig.~\ref{fig:compare_materials}(a) suggests a departure from the spin-spin interaction model although the discrepancy in the case of $\rm EuBa_{2}Cu_{3}O_6$ is less than that of $\rm La_2CuO_4$.\\


There are several reasons why one might expect deviations from our assumption that all of these materials
are described by a single-layer Heisenberg model.
The presence of single-ion anisotropies in $\rm K_2 Ni F_4$,
the presence of bilayers
with strong interlayer couplings in $\rm EuBa_2Cu_3O_6$, and the effect of temperature dependent
phonon-induced renormalizations of the antiferromagnetic exchange coupling, could all plausibly account
for the non-overlap of the data.
Empirically, we find that scaling the temperature by the N\'eel temperature, $T_N$, appears to provide a
good coincidence of the peak shift data on all three materials as shown in
Fig.~\ref{fig:compare_materials}(b), where $T_N$ replaces $JSz$ as the temperature scale.  That the data nearly collapse into one curve
is strongly suggestive of a spin-spin interaction mechanism for the energy shift of the RIXS peak,
which in turn hints at the magnetic nature of this peak.
One possible explanation of this data collapse upon using $T_N$ scaling is that
the 3D N\'eel temperature already takes into account various material specific properties mentioned above which affect the magnetic excitation spectrum.  This empirical observation needs to be tested in other insulating quantum magnets.

\subsection{Excitation peak position}
\label{subsect:disc_of_peakpos}

Although the temperature dependence appears to confirm that the 500 meV excitation shows characteristics of magnetic excitations, the question still remains as to the relatively weak nature of the magnon-magnon interaction within the two-magnon
excitation, as well as the precise value of the expected \textbf{q}=($\pi$ 0) peak position as determined in units of $J$, since not all calculations yield the same results (see Fig.~\ref{fig:mag_disp}(a)).  There are at least three possibilities for explaining the observed energy in the framework of magnons.  Sugai et. al. observed a distinct feature in the Raman spectrum of $\rm La_2CuO_4$ at $\sim$550 meV, which was attributed to a 4-magnon excitation.\cite{Sugai90}, or alternatively a phonon-induced magnon peak.\cite{Lee96,Lee97}  The matter is not resolved,\cite{Schmidt05n} but all scenarios for the Raman shoulder-feature involve multi-magnons, so much of the temperature and doping dependence analysis in this study should still apply.  However, the dispersion observed with RIXS would suggest that the excitation is a two-magnon (unless a 4-magnon has similar dispersion, but we cannot find such calculations in the literature).  We proceed below to explain the high energy, with the assumption that the RIXS excitation is indeed a two-magnon. \\

The 500 meV peak is only observable when the incident polarization is along the crystalline $c$-axis, leading to the suggestion in Ref.~\onlinecite{Hill08} that adjacent $ab$ planes could be involved, with the two magnons being created on adjacent planes, rather than on a single Cu-O sheet.  The small value of ${\rm J_{\perp}}$ would then explain the relatively weak interaction between the magnons.  Donkov and Chubukov also discussed the possibility that interplane interactions caused by the $c$-polarized photon could serve to weaken the magnon-magnon interaction.  Forte et al.\cite{Forte08} discussed another mechanism that might reduce the amount of energy renormalization at the zone-boundary.  They found that including the effect of longer range couplings (second and third nearest neighbors, as well as ring-exchange) increased the two-magnon
energy at finite \textbf{q} and their calculated energy is in agreement with this experiment, as seen in Fig. ~\ref{fig:mag_disp}(a).  Here, we present another possible resolution, based on a phase-space argument, which suggests that the interaction between in-plane magnons is a strong function of the total momentum ${\bf q}$.\\

\begin{figure}
\begin{center}
\epsfig{file=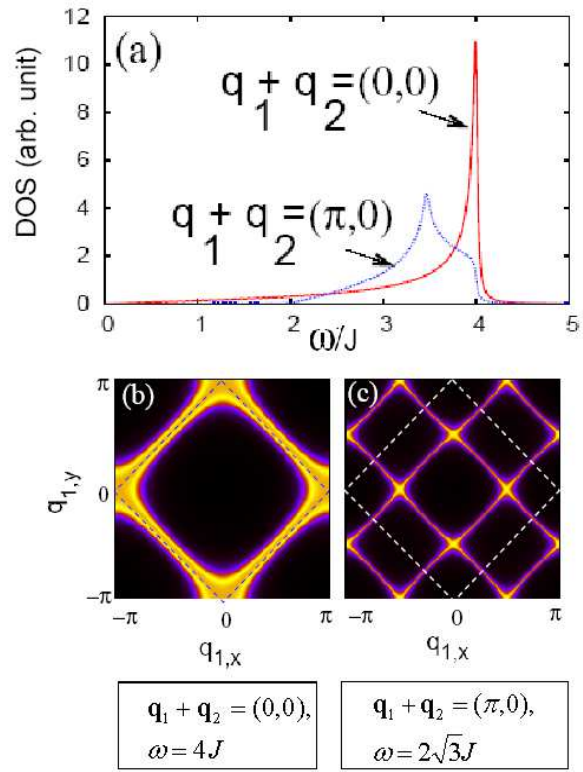,width=3.2in,keepaspectratio,}
\end{center}
\caption{(Color Online) (a): The two-magnon density of states for
noninteracting magnons with a total energy $\omega$ and two different
total momenta. (b): Intensity plot showing the single magnon states that contribute
to give a total momentum $\mathbf{q}_{tot}$=(0,0) at
$\omega$=$\omega_{peak}$=4\emph{J} (where the two-magnon density of states is a
maximum).  The two individual magnons have the maximum probability of
carrying momenta
$\mathbf{q}_2$=-$\mathbf{q}_1$=$\mathbf{k}$,
where $\mathbf{k}$ lies on the magnetic zone boundary. (c): Same as
in (b) but for $\mathbf{q}_{\mathbf{tot}}$=($\pi$,0) and
$\omega$=$\omega_{peak}$=2\emph{J}$\sqrt{3}$.  In this case, the
maximum probability is for individual magnons to have
$\mathbf{q}_1$=$\mathbf{q}_2$=$\mathbf{k}$,
where $\mathbf{k}$=($\pm$$\pi$/2,0), (0,$\pm$$\pi$/2).}
\label{fig:DOS}
\end{figure}

We begin by calculating the non-interacting two-magnon density of states (DOS). In Fig.~\ref{fig:DOS}(a), the
\textbf{q}-integrated DOS is plotted as a function of energy with the total
momentum $\mathbf{q}_1+\mathbf{q}_2$ fixed, where $\mathbf{q}_1$ and $\mathbf{q}_2$
denote individual momenta of the two constituent magnons. For the case of $\mathbf{q}_1$+$\mathbf{q}_2$=0, which is relevant for
two-magnon Raman scattering, the peak in the non-interacting two-magnon DOS occurs at an energy of 4$J$,
and it is well known that strong magnon-magnon interaction brings
this down to about 2.75$J$.\cite{Lyons88,Canali92,Chubukov95} However, the same DOS at $\mathbf{q}_1+\mathbf{q}_2$=$(\pi, 0)$
exhibits a broad peak at an energy of around 3.5$J$ (consistent with previous calculations, Ref. \onlinecite{Donkov07} for example) with a sharp
decrease in the DOS above 4$J$. Even with some uncertainty in the value of
$J$, the energy of the 500 meV peak observed at \textbf{q}=($\pi$ 0) is much closer to the non-interacting DOS peak, than is the 2-magnon Raman \textbf{q}=(0 0) peak to its corresponding DOS peak.\cite{energyscalenote} This naturally leads to the question of momentum dependence of magnon-magnon interactions within a 2-magnon.\\

To gain insight into the weakened interactions in the RIXS data, we have looked at the momenta of the component magnons, $\mathbf{q}_1$ and $\mathbf{q}_2$ for various total momenta ${\bf q}_1+{\bf q}_2$. Figure
\ref{fig:DOS}(b) shows constant energy cuts of the single magnon contributions to the two-magnon density of states at an energy of 4$J$ and a momentum of $\mathbf{q}_1+\mathbf{q}_2$=0.  The largest contributions come from single
magnons carrying the momentum shown as the bright color in this figure.
One can see that the zone-boundary magnons dominate the contributions to the peak in the two-magnon density of states. Because the magnon group velocity is zero at the
zone-boundary, there is ample time for these magnons to interact with one another.
Furthermore, there are a large number of
states corresponding to magnons with opposite momenta living on the zone boundary into which
scattering can occur.
As a result, interaction effects would be expected to be large at this total momentum - as is observed in the Raman experiments. In contrast, if we look at the single magnon contributions at a total momentum of
$\mathbf{q}_1+\mathbf{q}_2$=$(\pi, 0)$ and an energy of 3.5$J$, Fig. \ref{fig:DOS}(c), we find that the
individual magnons contributing to this peak are not concentrated
in the zone boundary region. For example, a pair of magnons near the
$(\pi/2, 0)$ position could add up to form a two-magnon mode with total
momentum $(\pi, 0)$.  Since these magnons can generally have a nonzero velocity with respect to each other,\cite{note2} we argue that the effective interaction between these two magnons is diminished, providing a natural qualitative explanation for the observed excitation energy.  We note that a similar idea was also proposed by Vernay et. al., who argued that magnon-magnon interaction is weakened because the average distance between spin-flips is larger when the total momentum is finite.\cite{Vernay07}\\

\section{Conclusion}
\label{sect:conclude}

We report a comprehensive investigation of the 500 meV peak in $\rm La_{2-x}Sr_{x}CuO_4$ observed with Cu $K$-edge resonant inelastic x-ray
scattering (RIXS).  We have carried out studies of incident energy, momentum, temperature, and doping dependence.  At temperatures below
300 K the peak energy shift of the RIXS peak is quantitatively consistent with both Raman\cite{Sugai88} and
neutron\cite{Coldea01} studies of $\rm La_2CuO_4$.  At high temperature the precipitous drop in the peak intensity is suggestive of magnon-phonon interaction.  However, the softening of the mode appears to be a result of coupling to spin fluctuations, as indicated by the scaling with $T_N$.  These quantitative and qualitative similarities with two-magnon Raman scattering, together with the observed dispersion and doping dependence, which are consistent with the expectations for a two-magnon excitation, provide compelling evidence for the magnetic nature of the 500 meV peak in La$_{2-x}$Sr$_x$CuO$_4$.\\

\begin{acknowledgements}
We would like to acknowledge Yong Cai, Alex Klotz, Tomohiro Sato and Harry Zhang for help with the experiments.  We also thank Clement Burns, Scott Coburn, Ercan Alp, Thomas Toellner, Harald Sinn, Ruben Khachatryan, Michael Wieczorek, Ayman Said, and Peter Siddons for the MERIX spectrometer design and implementation.  We also thank Alex Donkov and Luuk Ament for providing theoretical results.  The work at the University of Toronto was supported by the Natural Sciences and Engineering Research Council of Canada and the Early Researcher Award from the Ontario Ministry of Science and Technology.   AP acknowledges support from the Sloan Foundation.  The work at Brookhaven was supported by the U.S. DOE, Division of Materials Science, Contract No. DE-AC102-98CH10886; use of the APS was supported by the U.S. DOE, Basic Energy Sciences, Office of Science, under contract no. W-31-109-Eng-38;  RJB is supported at Lawrence Berkeley Laboratory by the office of Basic Energy Sciences, U.S. DOE under Contract No DE-AC03-76SF00098.  The scattering work at SPring-8 was performed under Common-Use Facility Programme of JAEA (Proposal No. A 2006A-EO4).
\end{acknowledgements}


\begin{thebibliography}{96}
\bibitem[{not()}]{note2}
\bibinfo{note}{When they are exactly at ($\pi$/2 0), the relative velocity is
  zero, but the weight of such cases would be small}.
\end{thebibliography}

\appendix

\section{Background Subtraction Methods}

Due to thermal and static diffuse scattering, there exists a large quasi-elastic ($\omega=0$) contribution to the background, and the tail of
this elastic background makes it difficult to study excitations below $\sim$500 meV.  For this reason, it is
important to subtract this quasi-elastic background to obtain the excitation spectrum.  We have employed three different methods to determine
the background, depending on experimental convenience or circumstance.  The first is to utilize the high-temperature spectrum as background.  This is most effective for studying temperature dependence, and is commonly used in many other spectroscopy experiments and is described in detail in Sec. \ref{subsect:temp}.  The other two methods are specific to RIXS, and it is worth discussing these in detail here.\\

\begin{figure}
\begin{center}
\epsfig{file=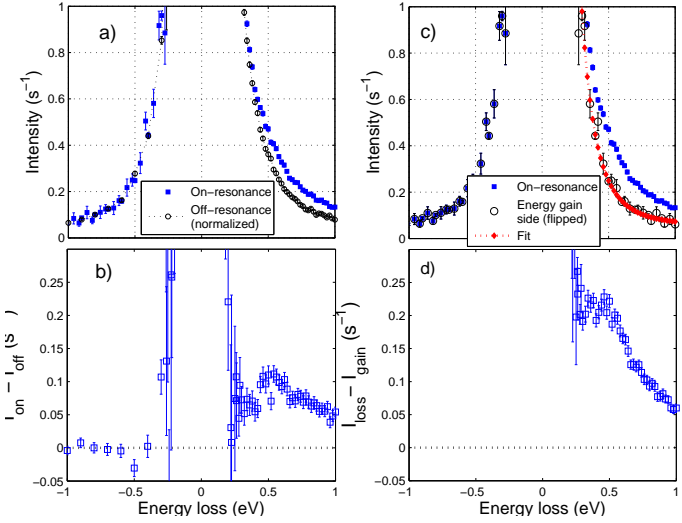,width=3.7in,keepaspectratio}
\end{center}
\caption{(Color Online) (a)  The RIXS intensity versus photon energy loss, for
$\rm La_{1.93}Sr_{0.07}CuO_4$ at \textbf{q}=$(\pi, 0)$ and T=20 K, for incident energies ``on-resonance'' $E_i$=8993 eV (filled squares)
and ``off-resonance'' (empty circles).  For improved statistics in the ``off-resonance'' spectrum we combined spectra of $E_i$=8981 eV
with $E_i$=8987 eV, whose low-energy spectra were identical within error bar.  The resultant spectrum was then normalized to match the
energy-gain side of the ``on-resonance'' spectrum.  (b) Subtraction of the spectra shown in (a), which has reasonable error bar only
above 0.3 eV energy loss. (c) The same ``on-resonance'' spectrum plotted as a function of energy loss (filled squares) and energy gain (open circles), and a fit of the energy gain spectrum to a Lorentzian lineshape plus constant (filled diamonds).  (d) Subtraction of a fit of the energy gain spectrum
from the original spectrum.}\label{fig:07_back_subtract}
\end{figure}

In the first of these, the ``off-resonance'' spectrum - measured with an incident photon energy well below the resonant energy - is subtracted from the ``on-resonance'' spectrum, obtained with the incident energy set to slightly above the absorption peak.  The off-resonance spectrum is normalized
such that the intensity of the quasi-elastic tail on the energy-gain side best matches with that of the on-resonant spectrum, as shown in
Fig.~\ref{fig:07_back_subtract}(a)-(b).  It is necessary to do so because the absorption changes dramatically in this energy range, resulting
in a large change of the quasi-elastic intensity.  While this method utilizes the unique resonance property of RIXS excitations, and eliminates artifacts from asymmetric analyzer tails, it can be time
consuming to obtain both on- and off-resonance scans with good statistics.\\

The final method relies solely on detailed balance.  Since we are measuring
excitations at a much higher energy than the thermal energy, the intensity of an excitation on the energy gain side
($\omega<0$) is essentially zero due to the principle of detailed balance.  Therefore, the energy gain spectra represents contributions purely from the resolution function and the
background.  Thus, subtracting a symmeterized version of the energy gain spectrum should remove both contributions, leaving just the inelastic processes.  This is accomplished by
reflecting the spectrum about $\omega=0~(\omega\rightarrow-\omega)$, and subtracting this energy gain spectrum from the original energy loss spectrum for $\omega>0$, as shown in
Fig.~\ref{fig:07_back_subtract}(c)-(d).  This method assumes that the resolution function (dominated by the analyzer) is symmetric.\\

Due to the steep tail of the elastic peak, the subtraction of the background spectrum by these
methods can be quite sensitive to even small energy shifts $\delta$$\omega$ of analyzer.  Small temperature variations in the experimental hutch, for example,
cause the lattice parameter of the analyzer to change, resulting in a 55~meV/$^{\circ}$C shift of the actual energy.  The largest
variations, of up to a couple of degrees $^{\circ}$C, can typically occur after the hutch is first closed.   To compensate for this effect, the
temperature near the analyzer was recorded for each data point and the energy values adjusted accordingly during the 9ID measurements.  Apart from the first several hours after the hutch is closed, the typical drift over the course of a scan was of the order of 20 meV or less, which mostly affected the part of the spectrum below 0.4 eV. \\

While these methods of estimating the contribution from the elastic tail should be equivalent in ideal conditions, a comparison of
Fig.~\ref{fig:07_back_subtract} (b) and (d) reveals that the energy-gain side subtraction method can result in an additional sloping background
when compared to the off-resonance method.  (It is important to note, however, that both methods give the same value for the peak position, the parameter of importance in this work.)  This might be due to non-resonant inelastic scattering
which would also be present in (and thus canceled by) the off-resonance scan.  Such non-resonant
scattering could be due to intraband excitations, and are thus expected to be more prevalent in doped samples.\\

In the course of these experiments, different circumstances favored one method or the other (for example, if the elastic peak happened not to be perfectly symmetrical
about $\omega$=0 due to analyzer imperfections, this would be problematic for the energy-gain side subtraction method). For the 9ID and BL11XU data we used off-resonance scans for background, while the energy-gain side subtraction method was employed in the 30ID momentum dependence scans.  It should be emphasized that the results from
using the different background subtraction methods were found to be consistent with each other and generally fall on the same curves for
temperature and momentum dependence (the one exception is discussed in \ref{subsect:Ei and q}).  \\






\end{document}